\documentclass[a4paper,fleqn,usenatbib]{mnras}

\usepackage{graphicx}	
\usepackage{amsmath}	
\usepackage{amssymb}	
\usepackage{lscape}
\usepackage{textcomp}
\usepackage{multirow}
\usepackage{times}

\title[]{Kinematic complexity around NGC~419: resolving the proper
  motion of the cluster, the Small Magellanic Cloud and the Magellanic
  Bridge}

\author[D. Massari]{
Davide Massari,$^{1,2}$\thanks{E-mail: davide.massari@inaf.it}
Silvia Raso,$^{1,3}$
Mattia Libralato,$^{4}$
Andrea Bellini$^{4}$
\\
$^{1}$INAF - Osservatorio di Astrofisica e Scienza dello Spazio di
Bologna, Via Gobetti 93/3, I-40129 Bologna, Italy\\
$^{2}$University of Groningen, Kapteyn Astronomical Institute, NL-9747
AD Groningen, The Netherlands\\
$^{3}$Dipartimento di Fisica e Astronomia, Universit\`{a} degli Studi
di Bologna, Via Gobetti 93/2, I-40129 Bologna, Italy\\
$^{4}$Space Telescope Science Institute, 3700 San Martin Drive,
Baltimore, MD 21218, USA 
}

\date{Accepted XXX. Received YYY; in original form ZZZ}

\pubyear{2020}

\begin{document}
\label{firstpage}
\pagerange{\pageref{firstpage}--\pageref{lastpage}}
\maketitle

\begin{abstract}

We present {\it Hubble Space Telescope} proper motions in the
direction of the star cluster NGC~419 in the Small Magellanic Cloud.
Because of the high precision of our measurements, for the first time
it is possible to resolve the complex kinematics of the stellar
populations located in the field, even along the tangential direction.
In fact, the proper motions we measured allow us to separate cluster
stars, which move on average with ($\mu_{\alpha}\cos\delta^{\rm NGC\,419}, 
\mu_{\delta}^{\rm NGC\,419}$) = ($+0.878\pm0.055$, $-1.246\pm0.048$) mas\,yr$^{-1}$, 
from those of the Small Magellanic Cloud and those belonging to a third kinematic
feature that we recognise as part of the Magellanic Bridge.  Resolving
such a kinematic complexity enables the construction of decontaminated
colour-magnitude diagrams, as well as the measurement of the absolute
proper motion of the three separate components. Our study therefore
sets the first steps towards the possibility of dynamically
investigating the Magellanic system by exploiting the resolved
kinematics of its stellar clusters.

\end{abstract}




\section{Introduction}

Star clusters are powerful tracers to investigate the formation and
the evolutionary history of their host galaxies; in this respect, the
study of their kinematics provides a wealth of information.  The most
immediate example is provided by the globular clusters (GCs) system of
the Milky Way. Thanks to the recent combined availability of proper
motion and spectroscopic radial velocity measurements, the orbits
around the Milky Way for almost all of the known GC have been
determined with great accuracy (\citealt{gaiahelmi18, vasiliev19,
  baumgardt19}). By building up on these measurements,
\cite{massari19} computed the GCs integrals of motion and were thus
able to associate them to known (\citealt{ibata94, helmi18,
  koppelman19, myeong19}) and yet-to-be-discovered merger events that
shaped the evolution of the Milky Way to its current structure
(\citealt{kruijssen2020}).

However, this kind of analysis is currently only feasible for the GC
system of our Galaxy. The reason is that other extra-Galactic GC
systems are too far away to be able to discern their proper motions
from that of the hosting galaxy. In fact, the typical velocity
difference between a globular cluster and the stars of the surrounding
environment is of the order of several tens of km\,s$^{-1}$, so that
the required proper motion precision to resolve such a difference is
of few tens of $\mu$as\,yr$^{-1}$ and increases linearly with the
distance. In the case of clusters embedded within the bulge of the
hosting galaxy the situation is even worse, as the required velocity
resolution drops to 5--10 km\,s$^{-1}$.  Yet, even with the limited
information provided by radial velocity alone, \cite{mackey19} were
able to exploit the kinematics of the GC system of M31 to conclude
that the galaxy likely experienced two significant merger
events. Adding the full three-dimensional kinematic information would
therefore enable a much more detailed reconstruction of the
evolutionary history of the galaxy.

One of the closest among other star cluster systems in the Local Group
is that of the Magellanic Clouds (MCs). Despite being small satellites
of the Milky Way (\citealt{vdm02, vdm14}), according to $\Lambda$CDM
cosmology they should still have experienced a complex history of
merging events (e.g., \citealt{sales13}), and such a prediction has
been recently supported by the dynamical confirmation that some of the
ultra-faint dwarf galaxies orbiting the Milky Way are actually
satellites of the Large Magellanic Cloud (LMC, \citealt{kalli18,
  erkal20, patel2020}). Moreover, the existence of the Magellanic
Bridge connecting the Large and the Small MCs (SMC) is a further
evidence of interactions between the two
(\citealt{zivick19}). Investigating the kinematics of the MCs star
clusters system could therefore help reconstructing the MCs past
history, yet resolving the clusters motion from that of the MC stellar
populations has so far proved to be beyond the capabilities of current
instrumentation like the {\it Gaia} mission.

One of the most effective ways to push the limits of astrometric
measurements beyond the current boundaries is to increase, whenever
possible, the temporal separation between different epochs of
observations of the same object. In this respect, the {\it Hubble
  Space Telescope} ({\it HST}) is often superior to {\it Gaia}, as the
latter is bound to scan the sky only for five years during the nominal
duration of the mission. This is why the most precise stellar proper
motions have been measured either by {\it HST} (\citealt{bellini18,
  libralato18, libralato19}) or by a combination of the two
space-based telescopes (\citealt{massari18, massari20}).  In this
paper, we combine a set of long temporal baseline {\it HST}
observations of the SMC cluster NGC~419 with the aim of resolving its
motion from that of the surrounding stars. The success in achieving
this objective could pave the road for the systematic investigation of
the resolved 3D kinematics of the star clusters system of the MCs.

The paper is organised as follows. Section \ref{data} presents the
dataset and the PM measurements. Section \ref{res} describes the
results we achieved on both NGC~419 and the SMC. The conclusions are
provided in Sect.\ref{concl}.

\section{Data analysis}\label{data}

\begin{figure}
	\includegraphics[width=\columnwidth]{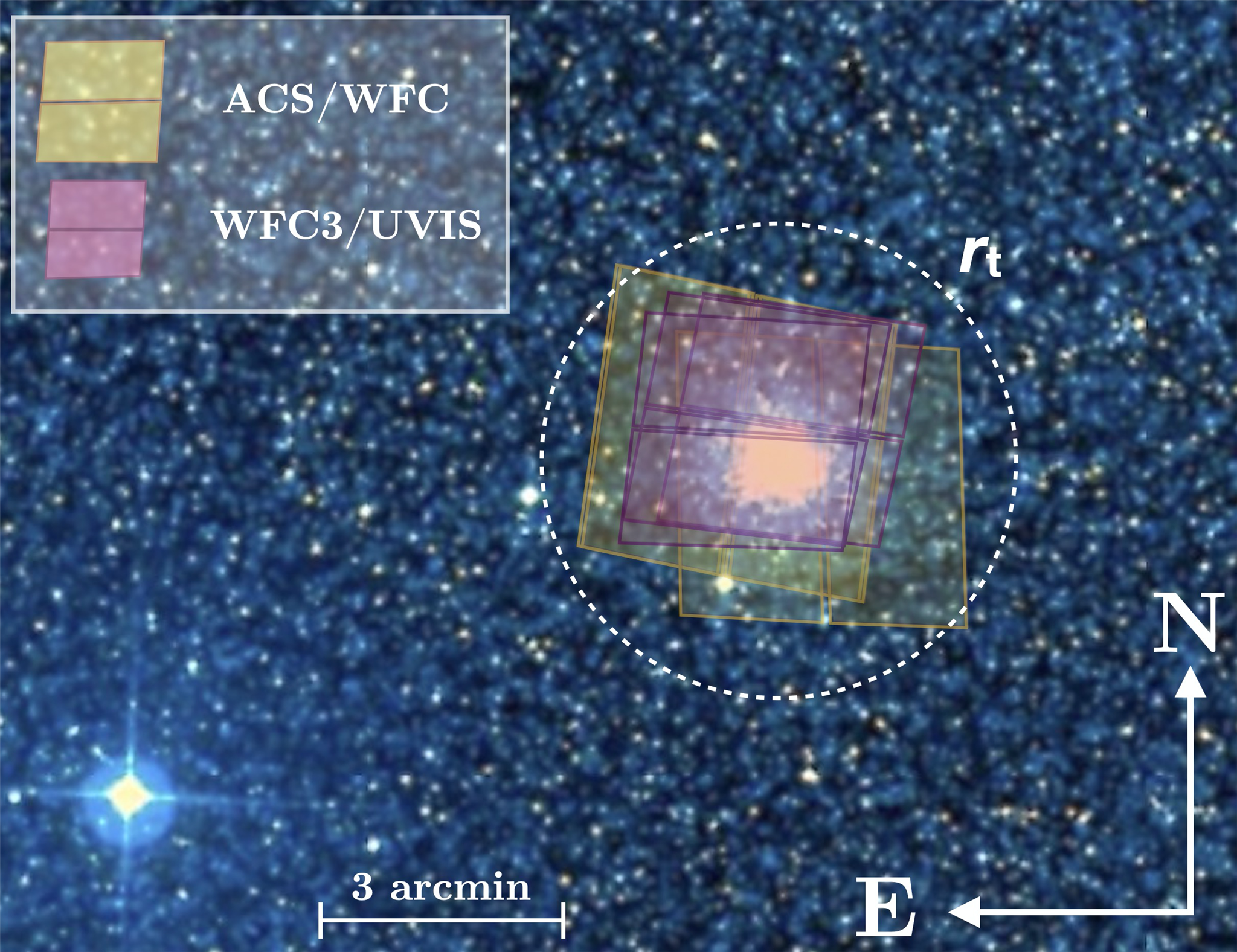}
        \caption{\small Sky distribution of the {\it HST} exposures used in the analysis.
        The yellow shaded areas represent observations taken with ACS/WFC, purple areas correspond to 
        WFC3/UVIS data. The angular scale of the image, orientation and tidal radius of the cluster (r$_{\rm t}\simeq174$ arcsec, \citealt{glatt09})
        are marked in white. The stellar field in the background has been taken from {\tt www.sky-map.org}.}\label{exp}
\end{figure}

We made use of the available \textit{HST} images of NGC~419 from the
Ultraviolet and Visible (UVIS) channel of the Wide Field Camera 3
(WFC3), obtained with the F336W and F438W filters, and from the Wide
Field Channel (WFC) of the Advanced Camera for Surveys (ACS), obtained with
the F555W and F814W filters. The list of the observations is given in
Table~\ref{tab:obs}, and the on-sky distrbution of each data set is 
shown in Fig.~\ref{exp}.

\begin{table*}[t!]
\caption{List of \textit{HST} images of NGC~419 used in this
  work.\label{tab:obs}} \centering {
\begin{tabular}{llllll}
\hline\hline
Program ID & PI & Epoch & Camera & Filter & Exposures \\
& & (yyyy/mm) & & & N $\times~t_{\mathrm{exp}}$ \\

\hline
GO-10396 & J. Gallagher & 2006/01     & ACS/WFC & F555W & $1\times 20\,{\rm s}$\\
& & & & F814W & $2\times 10\,{\rm s}$\\
& &  &  &  & $ 4\times 474\,{\rm s}$\\
& & 2006/07 & ACS/WFC & F555W  &  $ 2\times 20\,{\rm s}$ \\
& &  &  &  & $ 4\times 496\,{\rm s}$\\
& & & & F814W & $2\times 10\,{\rm s}$\\
& &  &  &  & $ 4\times 474\,{\rm s}$\\

\hline
GO-12257 & L. Girardi & 2011/08    & WFC3/UVIS & F336W & $1\times 400\,{\rm s}$\\
& & & & & $1\times 690\,{\rm s}$\\
& & & & & $2\times 700\,{\rm s}$\\
& & & & & $1\times 740\,{\rm s}$\\

\hline
GO-14069 & N. Bastian & 2016/08 & WFC3/UVIS & F438W & $1\times 70\,{\rm s}$\\
& & & & & $1\times 150\,{\rm s}$\\
& & & & & $1\times 350\,{\rm s}$\\
& & & & & $1\times 550\,{\rm s}$\\

\hline
GO-15061 & N. Bastian & 2018/09 & WFC3/UVIS & F336W & $2\times 1395\,{\rm s}$\\
& & & & & $1\times 3036\,{\rm s}$\\
& & & & F438W & $2\times 1454\,{\rm s}$\\

\hline
\end{tabular}}
\end{table*}

The photometric reduction was performed on the \texttt{\_flc} images,
which preserve the un-resampled pixel data for the stellar profile
fitting and are corrected to remove charge transfer efficiency (CTE,
see \citealt{jay10}), following the prescription given in
\citet{bellini17, bellini18}.  We refer to the aforementioned papers
for a detailed description of the photometric reduction procedure,
which we only briefly summarise in the following.  We first
created a list of bright objects in the field through a one-pass,
single finding procedure without neighbour subtraction, making use of
spatially variable library point-spread-function (PSF) of the
\textit{HST} detectors\footnote{Publicly available at
  \url{http://www.stsci.edu/~jayander/STDPSFs/}.}, which we fine-tuned
to each image using a set of bright, unsaturated and relatively
isolated stars. 
We corrected stellar positions for
geometric distortion with the solutions reported in \cite{jay06,
  bellini09, bellini11}.  Secondly, we used the software \texttt{KS2}
(see \citealt{bellini17} for details) for a multi-pass photometry.
This software is able to simultaneously perform the finding procedure
on all the images and to subtract neighbouring sources.  In this step,
the stellar positions in each exposure are transformed, by means of
six-parameter linear transformations, onto a common reference frame
system, based on the stellar positions in the Gaia Data Release 2
(DR2) catalogue (\citealt{gaia18}).  We calibrated the magnitudes to
the \texttt{VEGAMAG} photometric system as described in
\cite{bellini17} and \cite{raso19}, by adding to the instrumental
magnitudes the photometric zero point of the considered filter, and
the 3$\sigma$-clipped median difference between the aperture
photometry\footnote{The aperture photometry has been measured on the
  \texttt{\_drc} images with a 8-pixel aperture, and corrected for the
  finite aperture, see\\ \url{https://stsci.edu/hst/instrumentation/wfc3/data-analysis/photometric-calibration}.}  
  and the instrumental magnitudes.

\begin{figure}
	\includegraphics[width=\columnwidth]{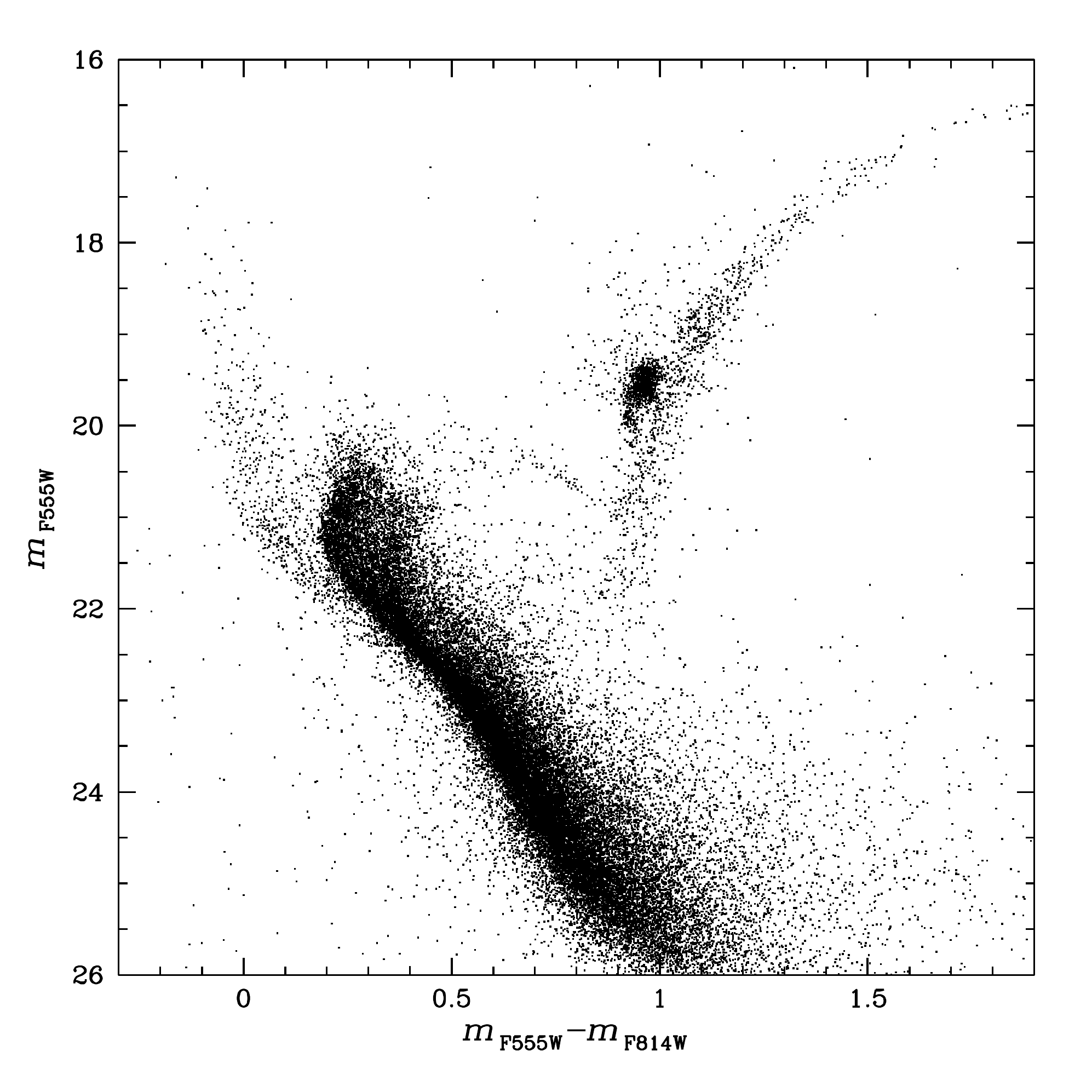}
        \caption{\small Optical $m_{\rm F555W}$ versus ($m_{\rm
            F555W}$-$m_{\rm F814W}$) CMD for all of the sources
          detected in the {\it HST} field surrounding the stellar
          cluster NGC~419.}\label{cmd}
\end{figure}

The calibrated colour-magnitude diagram (CMD) of NGC~419 is shown in
Fig.~\ref{cmd}.  The high photometric quality achieved allows us to
recognise the complexity of the field in terms of stellar
populations. The majority of the $44\,739$ measured stars is expected
to belong to the stellar cluster NGC~419, of which it is easily
possible to identify the extended main sequence turn-off (eMSTO) and
the sub-giant branch (SGB) regions.  The surrounding SMC is clearly
contaminating the CMD, as is evident from the young main sequence
populating its bluest region. Yet, the way the SMC contaminates the
cluster CMD is very difficult to describe precisely, because of the
intrinsic variety of its stellar populations. This is one of the
aspects that our proper motion analysis will help to solve.

After the photometric reduction, a number of stars were rejected as
poorly measured, based on several quality criteria. In particular, the
sources that survived this photometric selection (about 34\,000) were
those for which (i) the quality of the PSF fit described by the
\texttt{QFIT}\footnote{The \texttt{QFIT} value corresponds to the
  linear correlation coefficient between the pixel values and the PSF
  model.} parameter (\citealt{bellini17}) is better than the 95th
percentile of its distribution at any magnitude, (ii) the shape
parameter \texttt{RADSX} (\citealt{bedin08}) is lower than the 95th
percentile of its distribution at any magnitude, (iii) the fraction of
light from neighbouring sources within the fitting radius is not
larger than the light from the source itself.

\subsection{Relative proper motions}

Relative PMs were measured using the technique developed by
\cite{bellini14} and improved in \cite{bellini18} and
\cite{libralato18}.  A detailed description of the procedure, which we
only briefly summarise here, can be found in those papers.  The
procedure to measure the relative PMs is iterative. Each iteration
starts by cross-identifying stars in the raw catalogue of each
exposure with those on the master frame, once their positions have
been PM-shifted at the epoch of the raw catalogue. A six-parameter
linear transformation determined using a set of reference bright and
unsaturated cluster members is employed at this step. The use of
cluster members is the reason why the PMs are measured relative to the
bulk motion of the cluster, and thus have a zero mean value by
definition.  At the first iteration, stellar PMs are assumed to be
zero, so that cluster members are defined based on their location on
the CMD.  For each star, the master-frame transformed positions as a
function of epoch are fit with a least-squares straight line, the
slope of which is a direct estimate of the star's PM. This fitting
procedure is itself iterated after data rejection and sigma
clipping. The last least-squares fit is performed with
locally-transformed master-frame stellar positions, based on the
closest 45 reference stars, as this helps in correcting possible local
geometric distortion residuals and thus in mitigating small-scale
systematic effects.  At the end of each iteration, the stellar
cross-identification between the single exposures and master frames is
improved by adjusting the master frame positions to match the epoch of
each observation.  The iterative process converges when the difference
between master-frame positions from one iteration to the next are
negligible.  After convergence, we checked and corrected for spatially
variable and colour-dependent systematic effects as described in
\cite{bellini14}.

\begin{figure}
	\includegraphics[width=\columnwidth]{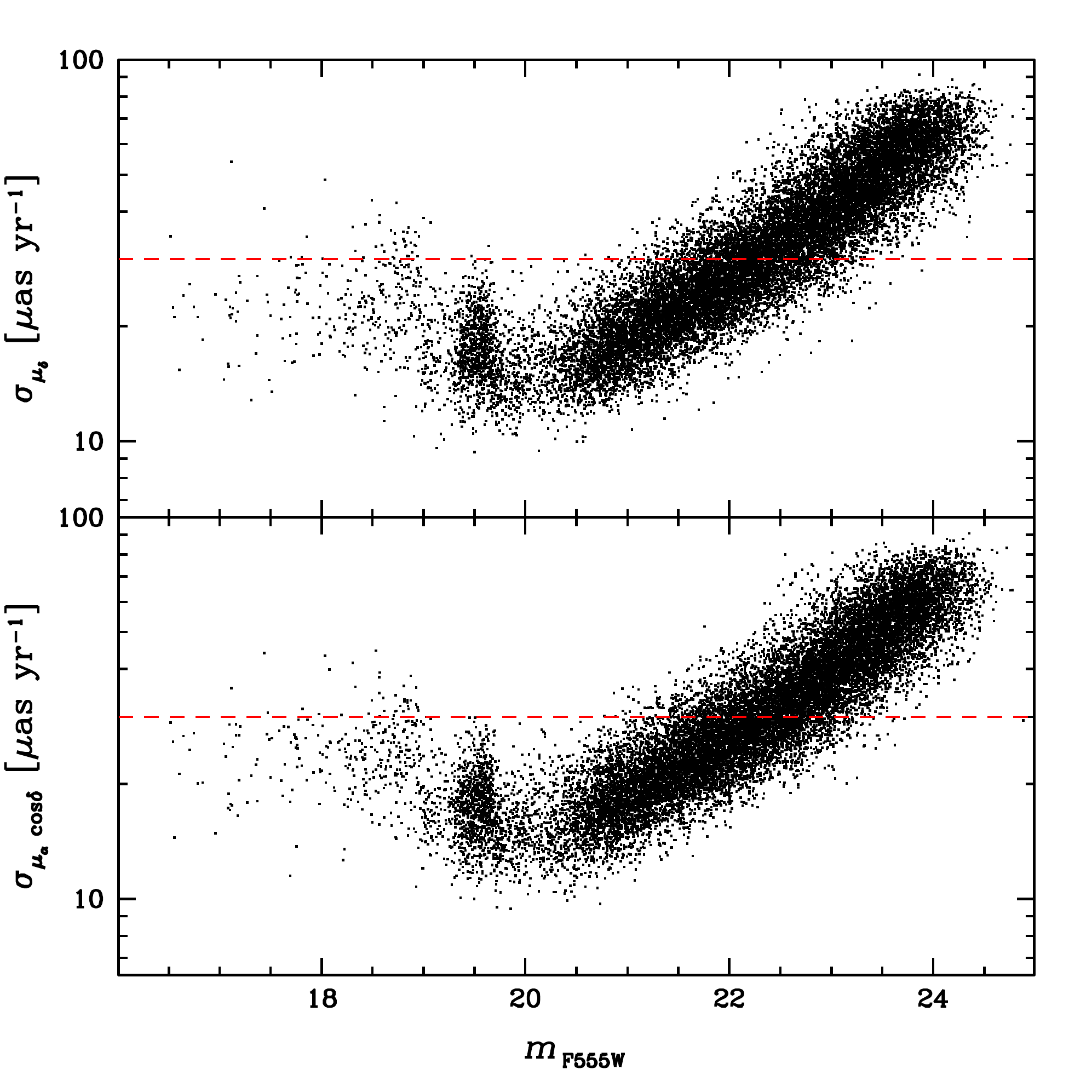}
        \caption{\small Distribution of PM errors as a function of the
          $m_{\rm F555W}$ magnitude for the sample of high-quality
          stars. The upper panel shows the PM error along Declination,
          while the lower panel shows the PM error along Right
          Ascension. The red dashed lines correspond to the cutoff 
          criterion applied to select the high-quality sample.}\label{pmerr}
\end{figure}

Astrometric quality criteria were also applied in order to only select
stars with reliable measurements. In particular, by following the
prescriptions given in \cite{libralato19}, we accepted stars for which
(i) the reduced $\chi^{2}$ of the PM fit is smaller than two in both
PM components, (ii) the fraction of positional measurements
effectively used for the PM fit is larger than 90\%, (iii) the error
on the PM is smaller than the 95th percentile of the its distribution
at any magnitude and (iv) the error on the PM is smaller than 0.1
mas\,yr$^{-1}$.  This leaved us with a total high-quality sample of
about 19\,000 stars. The behaviour of the PM uncertainty (see
\citealt{bellini18} for the details of its derivation) as a function
of the $m_{\rm F555W}$ magnitude for this sample is shown in
Fig.~\ref{pmerr}, where the best sources reach a PM precision of
$\sim$10 $\mu$as yr$^{-1}$.

\subsection{Absolute proper motions}

\begin{figure}
	\includegraphics[width=\columnwidth]{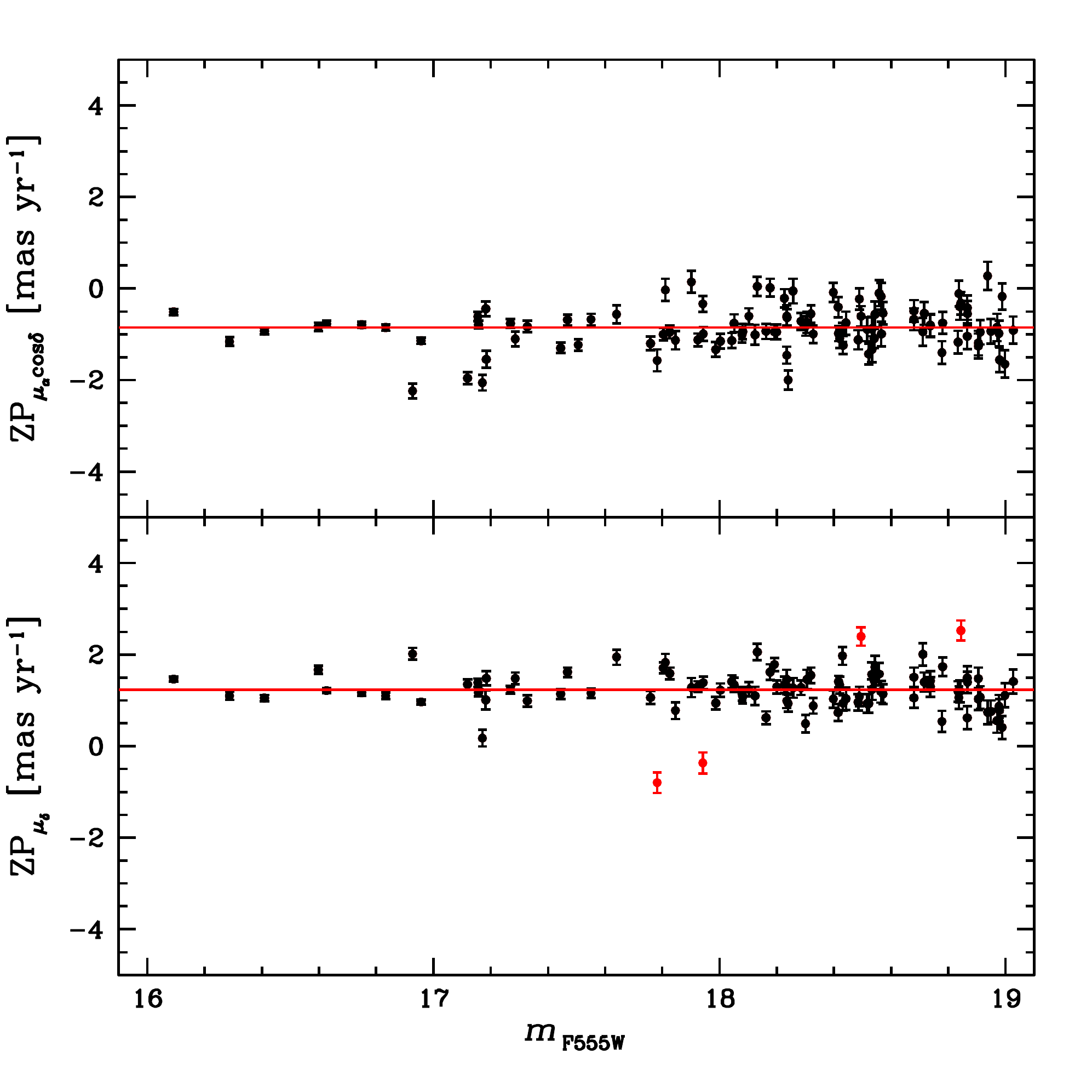}
        \caption{\small Absolute PM zero points for the two components
          computed from the stars in common between our high-quality
          sample and the {\it Gaia} catalogue. Red symbols indicate
          stars excluded by the adopted iterative
          3$\sigma$-clipping. The red solid lines mark the final mean
          zero points.}\label{zp}
\end{figure}

{\it HST} relative astrometry has been
historically brought to an absolute system using background galaxies
and quasars, since their absolute motions are known to be null due to
their large distance. The main limitation of this method is the low
number of this kind of faint objects in the typical {\it HST} field,
combined with shallow exposures times and stellar crowdining.  The
typical precision achieved on the absolute PM zero points (ZPs) using
background anchors varies from $\sim0.03$ (\citealt{sohn17}) to 0.1--0.2
mas yr$^{-1}$ (\citealt{massari13}, \citealt{libralato19}), depending on each
individual case. This limitation can in principle be overcome by using
Gaia stars as a reference, as they are bright in {\it HST} images and
significantly more numerous than suitable extragalactic calibrators, thus
offering more accurate and precise registration to an absolute
astrometric system (e.g., \citealt{libralato20}). 
For this reason, in this study we determined the
absolute PM zero points using the {\it Gaia} DR2
stars (\citealt{gaia18}) that are in common with our high-quality
sample.  The cross-match between the {\it Gaia} and our {\it HST}
catalogues has been performed by means of the CataPack suit of
software\footnote{\url{http://davide2.bo.astro.it/~paolo/Main/CataPack.html}}. The
adopted distance criterion for two stars to be matched is to have a
separation smaller than 0.4 arcsec (which corresponds to the effective
angular resolution of the {\it Gaia} DR2 observations, see
\citealt{lindegren18}) and a difference in magnitude smaller than one
mag.  In this way, we select 1755 stars having both {\it HST} and {\it
  Gaia} proper motions. Before estimating the absolute zero point,
though, we apply some further quality cuts in order to have the best
possible reference sample. In particular, we excluded all of the stars
with renormalised unit weight error (ruwe) $>1.4$
(\citealt{lindegren18}), stars with {\it Gaia} proper motion
uncertainty larger than 0.3 mas\,yr$^{-1}$ and stars within a distance
of 50 arcsec from the cluster centre, where crowding is more severe
for {\it Gaia}. This leaves us with a total of 105 reference stars.
The application of the 0.1 mas\,yr$^{-1}$ PM error cut adopted for
{\it HST} stars would have included only 24 stars, and this is why we
chose a more generous cut for the {\it Gaia} stars.  The difference
between the selected {\it HST} and {\it Gaia} proper motions is shown
in Fig.~\ref{zp}, where the error bars represent the associated 
$1\sigma$ uncertainties, and each PM component is plotted separately.  In
order to determine the absolute zero points (ZPs), we finally applied an
iterative 3$\sigma$-clipping algorithm (rejected stars are shown in
red in Fig.~\ref{zp}) that provided the following
solution:\\ 
ZP$_{\rm \mu_{\alpha}\cos\delta} = -0.878\pm0.055$ mas\,yr$^{-1}$\\ 
ZP$_{\rm \mu_{\delta}} = +1.226\pm0.048$ mas\,yr$^{-1}$,\\ 
where the quoted uncertainties are the sum in quadrature of two
terms. The first term is given by the standard error of the
mean value of the two distributions, while the second term describes
the {\it Gaia} systematic error on its PMs, and amounts to 0.03 mas\,yr$^{-1}$
(see \citealt{gaiahelmi18}). Neither of the two proper motion
zero points shows a trend with stellar magnitude, further supporting
the quality of our selection.

Parallax effects on the measured astrometric shifts among the different 
epochs should also be quantified, as they are a source of systematic errors in our analysis. 
The mean {\it Gaia} parallax of the stars used to determine the ZPs is $\pi_{Gaia}=0.07$ 
mas, while for the sources located in the SMC we can assume a typical parallax 
$\pi_{\rm SMC}=0.017$ mas. By following the prescriptions given in \cite{massari13},
such a difference in parallax translates to a maximum apparent PM of
0.008 mas\,yr$^{-1}$, which is negligible when compared to the uncertainty
on the absolute PM ZPs.

\begin{figure}
	\includegraphics[width=\columnwidth]{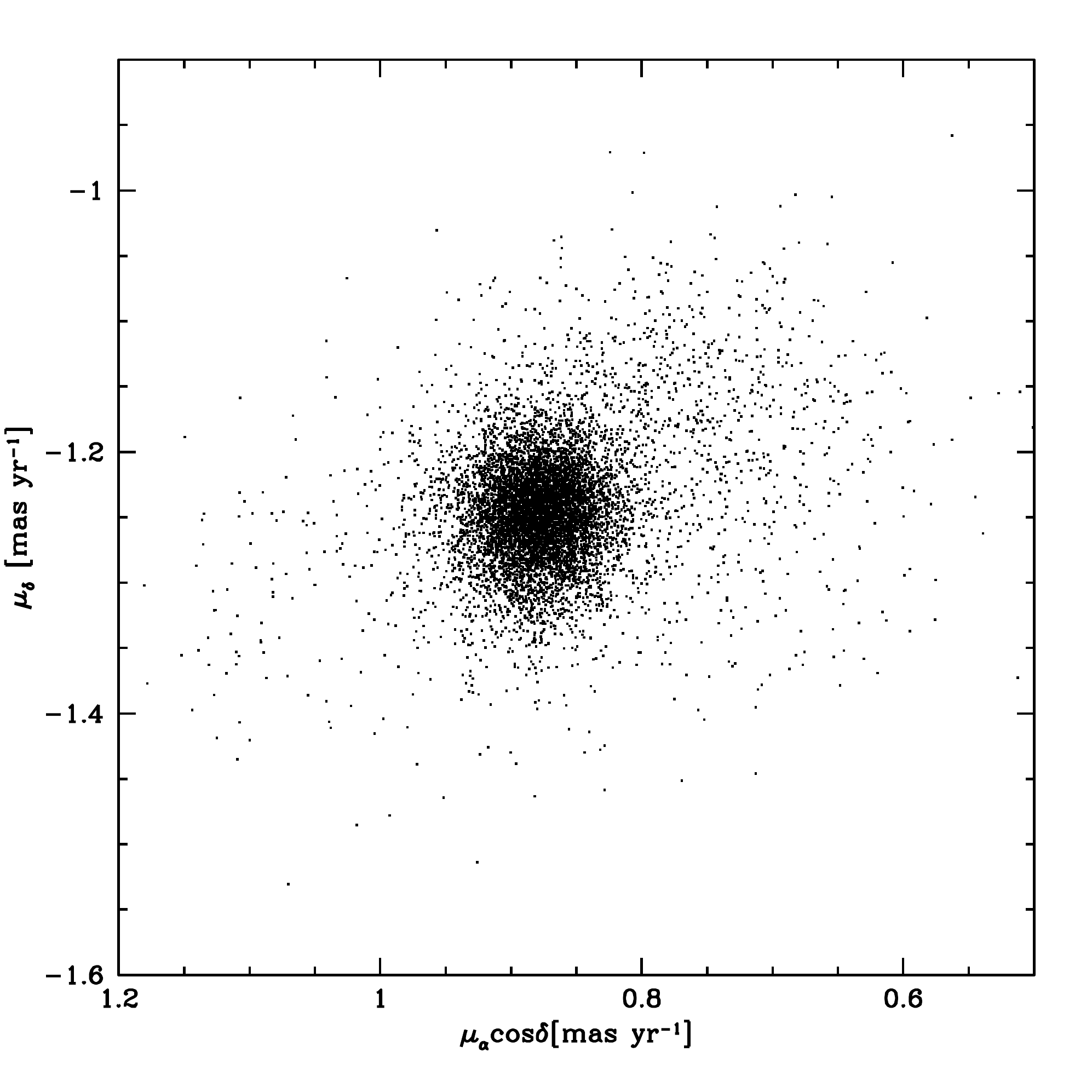}
        \caption{\small VPD for the stars belonging to the
          high-quality sample and with a PM error smaller than 30
          $\mu$as\,yr$^{-1}$ in both components.}\label{vpd_all}
\end{figure}

The vector point diagram (VPD) resulting from the application of the
zero points is shown in Fig.\ref{vpd_all}. A further cut on the PM
error, which we require to be smaller than 30 $\mu$as\,yr$^{-1}$ in
both components (see the dashed red lines in Fig.~\ref{pmerr}), 
has been applied in order to highlight the most important kinematical 
features, and will be maintained throughout the rest of the analysis 
unless stated differently.

\section{Population selections and results}\label{res}

The aim of our analysis is to distinguish NGC~419 members from the
surrounding SMC stars by exploiting their kinematics.  In order to do
so we require two clean samples of stars belonging to the two
different populations.  We remark that we do not care about how
complete these samples are at this stage, but rather that the
contamination from other populations is as little as possible.

\subsection{NGC~419}

\begin{figure*}
	\includegraphics[width=\columnwidth]{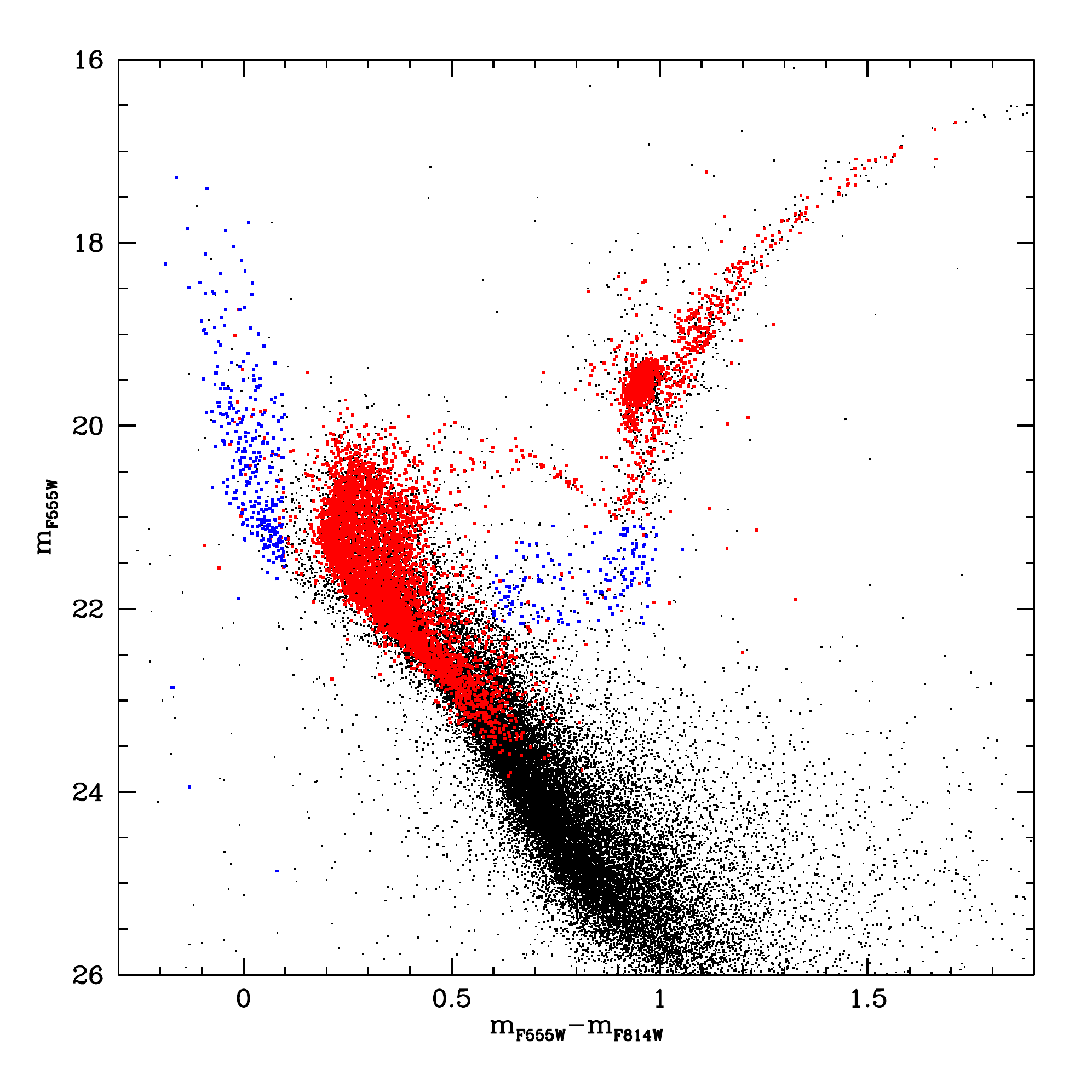}
	\includegraphics[width=\columnwidth]{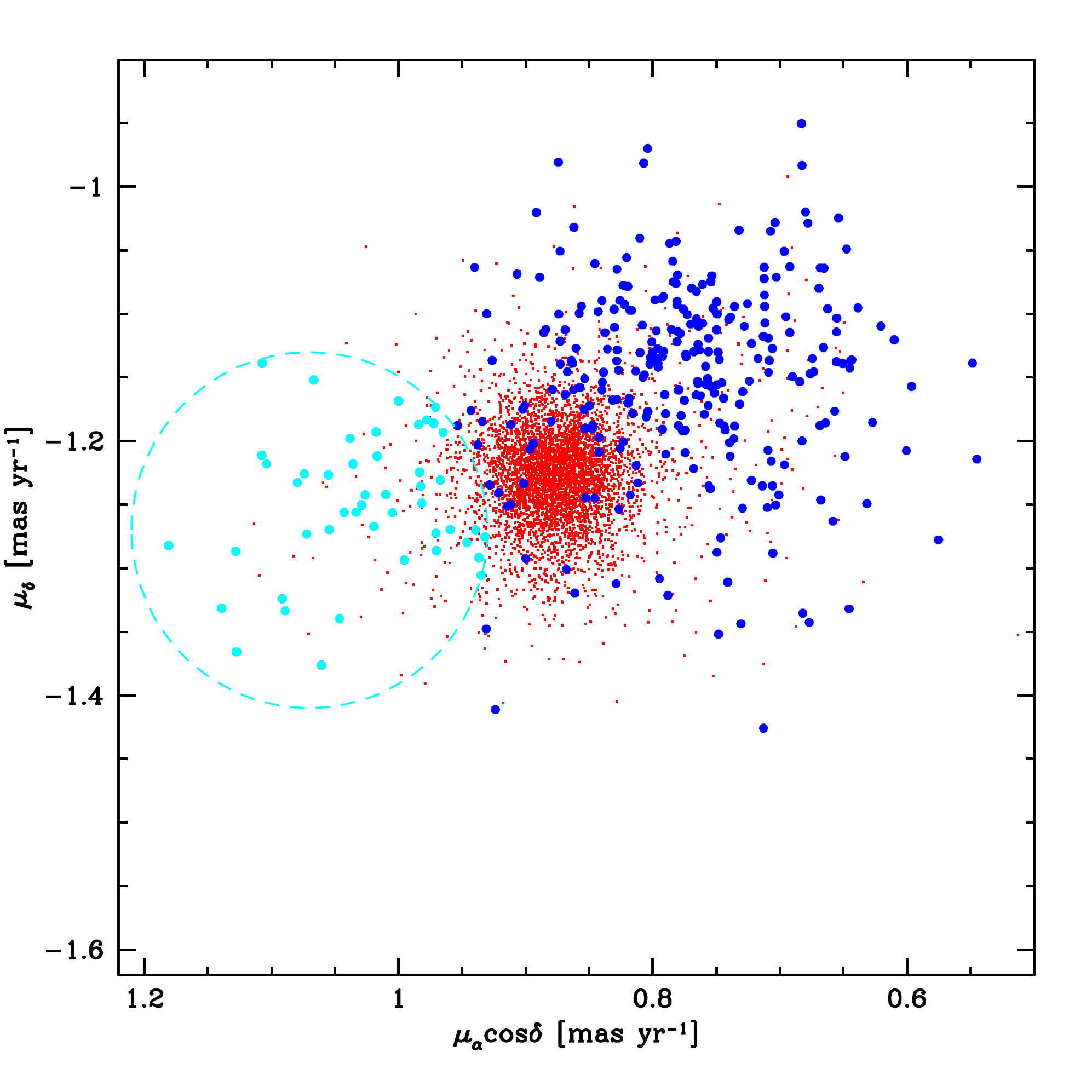}
        \caption{\small {\it Left panel:} distribution in the CMD of
          the clean samples of cluster members (red symbols) and SMC
          field stars (blue symbols).  {\it Right panel:} distribution
          of the same samples in the VPD. SMC field stars split into
          two separated features, one trailing and one ahead of the
          GC. The latter is highlighted with cyan
          symbols.}\label{cmd_sel}
\end{figure*}

To select stars belonging to the star cluster, we adopted the
following criteria:
\begin{description}
\item{$i$) we only considered stars closer than 40 arcsec from NGC~419
  centre, as the density of its members is obviously higher in the
  innermost regions;}
\item{$ii$) we selected stars within 80 $\mu$as\,yr$^{-1}$ from the
  mean motion of the bulk of sources in the VPD, which is at (0,0)
  mas\,yr$^{-1}$ of the relative proper motion VPD, by construction.}
\end{description}
The location on the CMD of this sample of NGC~419 stars is shown with
red dots in the left panel of Fig.~\ref{cmd_sel}.  NGC~419 is one of
the first star clusters where an extended main sequence turn off has
been detected (\citealt{glatt08}). Our selection leads to a CMD where
this feature and the fainter main sequence are defined remarkably
well, as well as the more evolved giant branches.  Another peculiar
photometric feature of NGC~419 is the so-called secondary red clump,
which according to \cite{girardi09} is due to the simultaneous
presence in the cluster of stars that burn Helium in degenerate and
non-degenerate cores. These authors had already shown that the
secondary red-clump is likely a genuine feature of the cluster, rather
than made up of SMC stars, based on statistical arguments. Thanks to
our proper motion analysis, here we can firmly confirm that the
feature is in fact described by cluster members.  We underline that, to
our knowledge, this is the first kinematically decontaminated CMD of a
GC in the MC ever presented in the literature. Even though the focus
of this paper is the kinematics of the cluster and the SMC, our
analysis shows that the proper motion-based decontamination of CMDs of
MC star clusters is now within the reach of {\it HST}.

Once a clean sample of cluster stars has been defined, it is possible
to investigate its distribution in the VPD. This is shown in the right
panel of Fig.~\ref{cmd_sel}.  The members of NGC~419 (red symbols)
clearly populate the bulk of the distribution. Since this is centred
on the origin of the VPD of the relative proper motions by construction, the
absolute proper motion of NGC~419 is given by the absolute proper
motions zero points, with opposite sign:\\ 
$\mu_{\alpha}\cos\delta^{\rm NGC\,419} = +0.878\pm0.055$ mas
\,yr$^{-1}$\\
$ \mu_{\delta}^{\rm NGC\,419} = -1.226\pm0.048$ mas\,yr$^{-1}$.

Coupled with the cluster position (\citealt{glatt08}), distance
($\sim59$ kpc, \citealt{goud14}), and radial velocity of $v_{\rm
  sys}=190.5$ km\,s$^{-1}$ (\citealt{kamann18}), our proper motion
estimate enables future investigations of the dynamics of this system
within the complex gravitational potential of the MCs.

\subsection{SMC}

A clean sample of SMC stars has to be selected among the sources that
are not labelled as NGC~419 members. Given that the SMC stellar
population is less numerous and has a larger velocity dispersion, a
selection on the VPD is more subtle, as several SMC stars would end up
within the limit of the cluster selection. For this reason, this time
we preferred to exploit the CMD, and defined the following criteria:
\begin{description}
\item{$i$) we selected stars more distant than 40 arcsec from NGC~419
  centre;}
\item{$ii$) we applied a colour cut such to select stars along the
  young main sequence, nominally ($m_{\rm F555W}-m_{\rm F814W})<0.1$;}
\item{$iii$) we selected a region in the CMD that includes the faint
  and old SGB, at $0.6<(m_{\rm F555W}-m_{\rm F814W})<1.1$ and
  $21<m_{\rm F555W}<22.2$.}
\end{description}
The location of the SMC sample in shown in the left panel of
Fig.~\ref{cmd_sel} as blue points.

When looking at selected SMC stars in the VPD, two separate features
stand out clearly. The first and more populated one is a broad clump of
stars that trails NGC~419 along Right Ascension (R.A.), and is located
at about ($\mu_{\alpha}\cos\delta$, $\mu_{\delta}$)$=(0.8,-1.15)$
mas\,yr$^{-1}$. The second feature is instead less populated and moves faster
than the cluster along R.A. It is highlighted with cyan symbols in the
right panel of Fig.~\ref{cmd_sel} for sake of clarity.

\begin{figure}
	\includegraphics[width=\columnwidth]{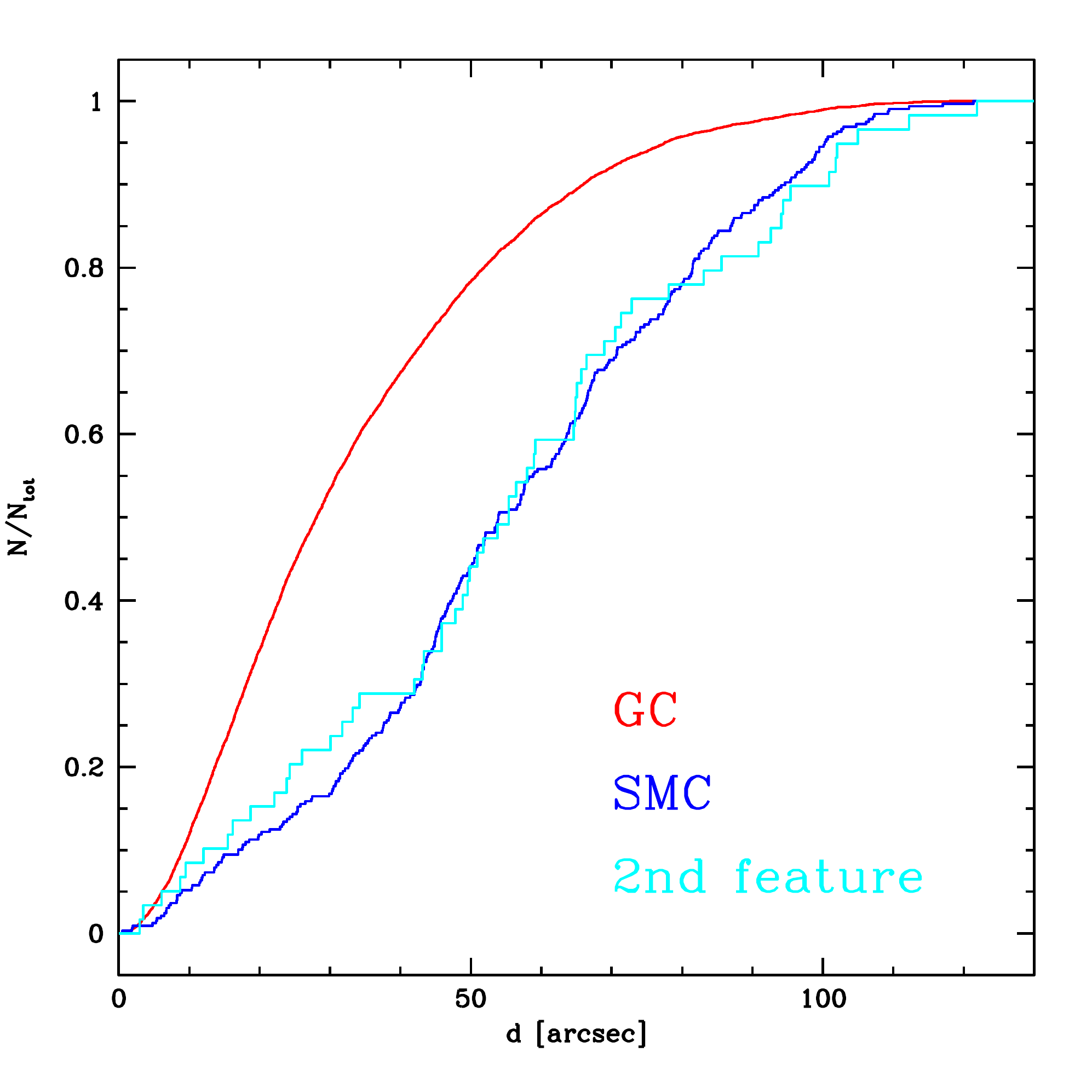}
        \caption{\small Radial distributions of cluster members (red
          line), and the two sub-populations of SMC field stars (blue
          and cyan symbols).}\label{rad}
\end{figure}

In order to better interpret the nature of these two features, we first
investigated their spatial distribution. As shown in Fig.~\ref{rad},
both are less centrally concentrated than the cluster, as expected for
field populations. Moreover, their radial distribution is basically
indistinguishable.  A Kolmogorov-Smirnov test run on the two samples
confirms that the probability for them to have been extracted from two
populations sharing the same distribution on the sky is of 96.2\%.

\begin{figure}
	\includegraphics[width=\columnwidth]{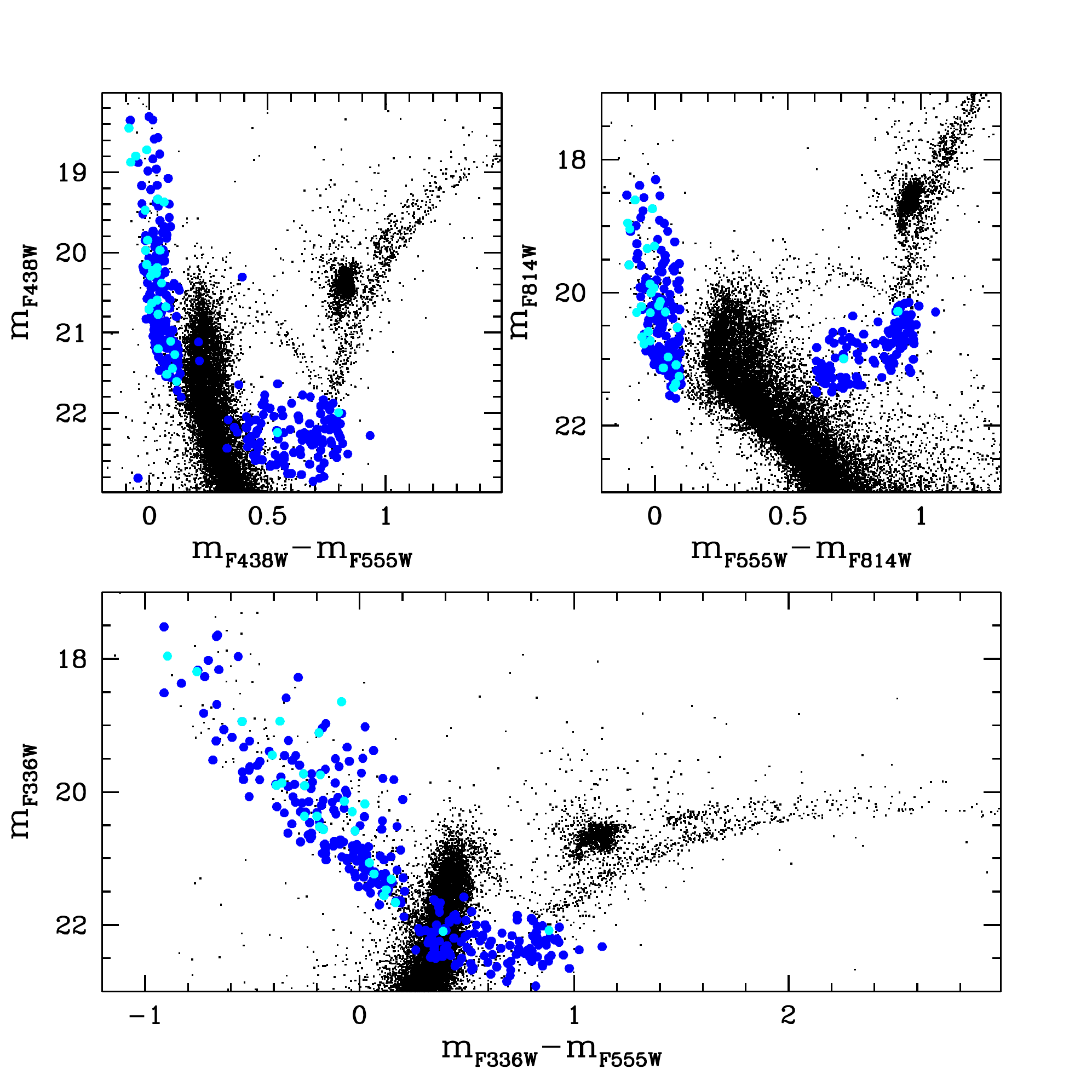}
        \caption{\small CMDs in different {\it HST} bands showing how
          the main SMC sub-population (blue symbols) is made up of
          both young and old stars, while the secondary field
          sub-population (cyan symbols) mainly includes young
          stars.}\label{blob}
\end{figure}

On the other hand, a further inspection of the CMD revealed an
important difference. As shown in Fig.~\ref{blob}, the stars belonging
to the main feature (blue points) populate both the young main
sequence and the old SGB, regardless of the adopted combination of
filters. Instead, the secondary feature is predominantly made up of young main
sequence stars (cyan dots).  We therefore recognise the main feature
as the one describing the stellar component of the main body of the SMC. 
Its 3$\sigma$-clipped mean absolute proper motion amounts to\\ 
$\mu_{\alpha}\cos\delta^{\rm SMC} = +0.777\pm0.005\pm0.055$ mas\,yr$^{-1}$\\
$\mu_{\delta}^{\rm SMC} = -1.141\pm0.003\pm0.048$ mas\,yr$^{-1}$,\\
where the second term of the associated uncertainty represents the
error on the absolute proper motion zero points.  Our field lies about
1 degree off of the nominal centre of the SMC, which is moving away
from us with a radial velocity of 145.6 km\,s$^{-1}$
(\citealt{harris06}), at a distance of 63 kpc
(\citealt{cioni2000}). Prior to compare our measurements with others
available in the literature, we must therefore correct for the
perspective effect introduced by the different lines of sight. By
using the formalism described in \cite{vandeven2006} and
\cite{gaiahelmi18}, we obtain an estimate for the PM of the SMC centre
of mass (COM) of:\\ 
$ \mu_{\alpha}\cos\delta^{\rm SMC,COM} =+0.711\pm0.004\pm0.055$ mas\,yr$^{-1}$\\
$ \mu_{\delta}^{\rm  SMC,COM} = -1.190\pm0.003\pm0.048$ mas\,yr$^{-1}$.

As a comparison, \cite{kalli13} estimated the proper motion of the SMC
centre of mass to be
$(\mu_{\alpha}\cos\delta,\mu_{\delta})=(0.772\pm0.063,-1.117\pm0.061)$
mas\,yr$^{-1}$, whereas \cite{gaiahelmi18} estimated the mean PM of
the SMC to be $(\mu_{\alpha}\cos\delta,\mu_{\delta})=(0.797\pm0.030,
-1.220\pm0.030)$ mas\,yr$^{-1}$.  Both the results are in good agreement 
with our estimate.

The high quality of our PMs therefore allows us to clearly separate
the motion of NGC~419 from that of the SMC. The net PM difference
amounts to $\Delta\mu_{\alpha}\cos\delta=-0.1\pm0.005$ mas\,yr$^{-1}$
along R.A. and to $\Delta\mu_{\delta}=0.085\pm0.003$ mas\,yr$^{-1}$
along Declination, which at the distance of the SMC (${\rm d}_{\rm
  SMC}\sim60$ kpc, \citealt{cioni2000, muraveva18}) translates to an
overall velocity difference of about $37.3\pm1.4$ km\,s$^{-1}$. This
value exceeds the velocity dispersion that has been measured for the
SMC from a sample of 2046 red giant stars by \cite{harris06}, who
found v$_{\rm disp}=27.5\pm0.5$ km\,s$^{-1}$. We cannot draw any clear
conclusion on whether or not such a difference is significant enough
to advocate for a peculiar origin for the cluster.  However given the
young cluster age ($\sim1.5$ Gyr, \citealt{glatt08}) and the lack of
information on the anisotropy of SMC stars, we believe that our
results reasonably support an in-situ origin for NGC~419 within the
SMC.

The secondary feature is thus made up of young stars, which are
homogeneously distributed across the FoV as a field population. To
verify whether these stars could belong to the Milky Way disk, we
analysed their {\it Gaia} parallaxes and found out that they are too
small (the median value is $\sim0.02$ mas) to be foreground
contaminants (for which one should expect parallax values about one
order of magnitude larger).  The crucial piece of information to
interpret this feature is ultimately given by the location of NGC~419
itself.  Its coordinates lies at about 1 degree from the centre of the
SMC and along the direction of the Magellanic Bridge
(\citealt{hindman63}).  The Magellanic Bridge is a gaseous structure
connecting the two MCs that also hosts stars. The stellar component of
the Bridge is mostly composed of young stars (\citealt{irwin85}),
though evidence for the presence of stars older than 1 Gyr also exists
(e.g., \citealt{bagheri13, belokurov17}).  The PM of stars in the
Magellanic Bridge has been very recently measured by \cite{zivick19} and
\cite{schmidt20}, and have lent further support to the hypothesis that
this structure has formed from the past mutual interaction of the MCs
(\citealt{besla12}).  In \cite{schmidt20}, in particular, the PM
measurements extend from the LMC up to a distance of $\sim4$ degrees
from the SMC centre. The field around NGC~419 is even closer to the
centre of the hosting galaxy, though, and this prevents a direct
comparison with previous estimates.  For the sample of stars belonging
to the secondary feature of the VPD we measure a mean absolute proper
motion of\\ 
$ \mu_{\alpha}\cos\delta^{\rm Bridge}=+1.025\pm0.009\pm0.055$ mas\,yr$^{-1}$\\
$ \mu_{\delta}^{\rm Bridge} = -1.247\pm0.007\pm0.048$ mas \,yr$^{-1}$.

These values look consistent with the extrapolation towards smaller
distances from the SMC centre of the N-body models presented by
\cite{schmidt20}, shown in their Fig.~10.  When transformed to an
orthografic projection centred on the SMC using the equations given in
\cite{gaiahelmi18}, the PM of our secondary feature become $(\mu_{\rm
  x},\mu_{\rm y})=(0.210\pm0.055,-0.100\pm0.048)$
mas\,yr$^{-1}$. These values are also in reasonable agreement with the
models described in \cite{zivick19}.

\begin{figure}
	\includegraphics[width=\columnwidth]{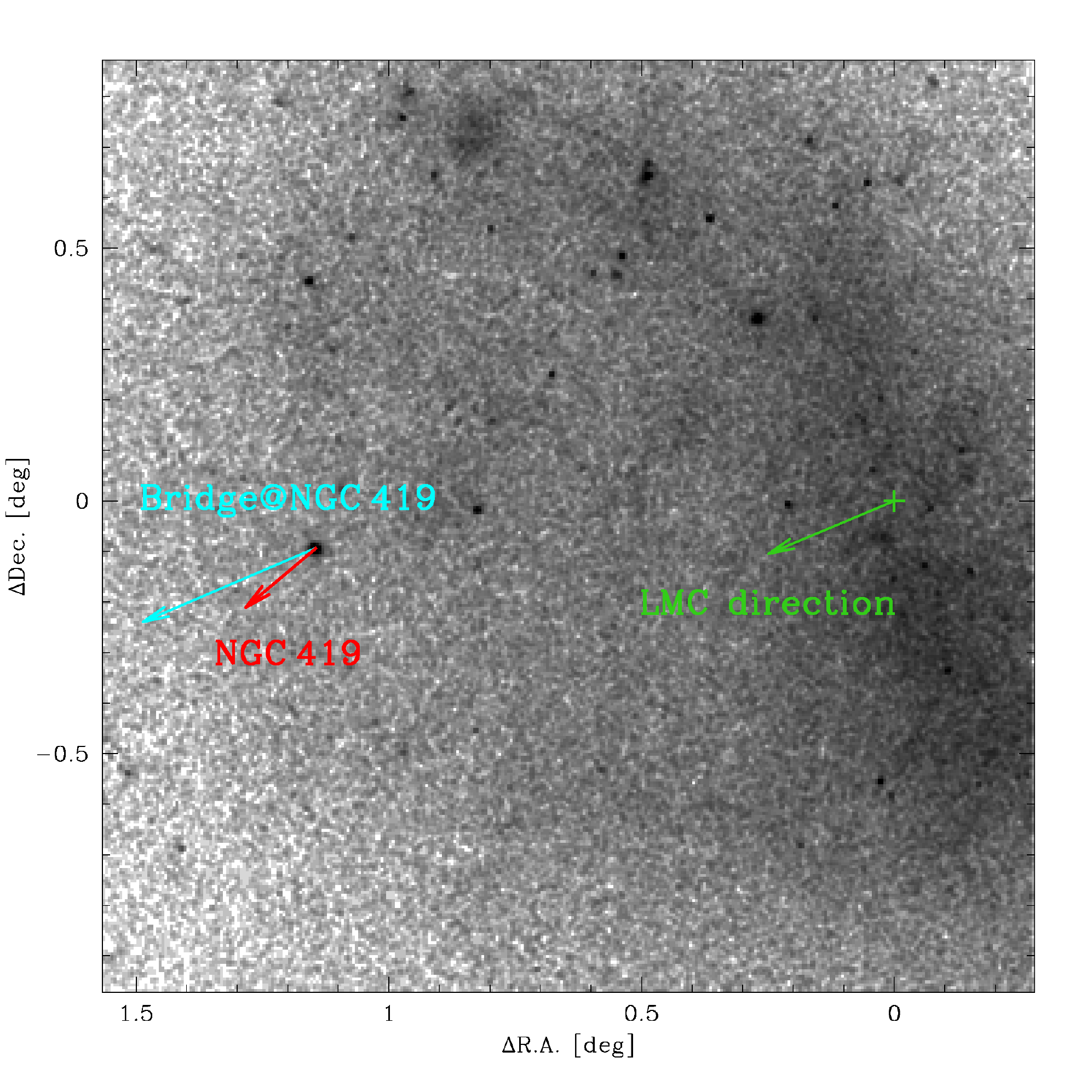}
        \caption{\small {\it Gaia} DR2 density map of the SMC showing
          the location of our field of view and the motion of NGC~419
          and Magellanic Bridge stars in a reference system where the
          SMC is at rest. The direction connecting the SMC and the LMC
          centres is also shown as a green arrow.}\label{map}
\end{figure}

Therefore, after putting together this kinematical information with
the spatial and age-related arguments, we interpret the secondary
feature of the VPD as made up of stars belonging to the Magellanic
Bridge in the proximity of the SMC centre. Figure \ref{map} summarises
our findings by showing how NGC~419 (red arrow) and the Bridge stars
(cyan arrow) move in a reference system centred on the SMC (marked
with a green cross, the underlying density map is based on {\it Gaia}
DR2 data and only includes sources brighter than $G=19$) and where 
the SMC is at rest.  If our interpretation is correct, then our 
measurements would provide a useful constraint on the overall kinematics 
of the Bridge in a region that has so far been unexplored.

\section{Summary and Conclusions}\label{concl}

We have analysed multi-epochs {\it HST} observations of the SMC
globular cluster NGC~419 with the aim of measuring the PM of stars in
its field.  Our dataset span a temporal baseline of 12.63 years and,
coupled with the exquisite astrometric capabilities of the telescope,
allowed us to achieve proper motion measurements as precise as 10
$\mu$as\,yr$^{-1}$.  With such a precision, we were able to resolve the
intricate kinematics of the stellar populations sampled by our
observations. In particular:
\begin{description}
\item{i) NGC~419 members were efficiently isolated from the field
  contaminants. The resulting CMD of the cluster confirmed the
  existence of peculiar features such as the eMSTO and the secondary
  red clump, which for the first time are decontaminated from
  non-member stars on an individual (and not statistical) basis;}
\item{ii) the presence of stars with a {\it Gaia} DR2 PM
  measurements in our field allowed us to determine the absolute PM of
  NGC~419, which is $\mu_{\alpha}\cos\delta^{\rm NGC\,419} =
  +0.878\pm0.055$ mas\,yr$^{-1}$, $\mu_{\delta}^{\rm NGC\,419} =
  -1.226\pm0.048$ mas\,yr$^{-1}$;}
\item{iii) field stars clearly describe a main, homogeneous feature
  in the VPD, populated by both young and old stars. We associated
  this feature to stellar component of the main body of the SMC, and 
  determine its perspective corrected absolute proper motion to be
  $\mu_{\alpha}\cos\delta^{\rm SMC,COM} = +0.711\pm0.004\pm0.055$
  mas\,yr$^{-1}$, $\mu_{\delta}^{\rm SMC,COM} = -1.190\pm0.003\pm0.048$
  mas\,yr$^{-1}$, in excellent agreement with previous measurements;}
\item{iv) a few field stars clump around a secondary feature of the
  VPD that is solely populated by young stars, and which bulk motion
  is $\mu_{\alpha}\cos\delta^{\rm Bridge} = +1.025\pm0.009\pm0.055$
  mas\,yr$^{-1}$, $\mu_{\delta}^{\rm Bridge} = -1.247\pm0.007\pm0.048$
  mas\,yr$^{-1}$. Based on the location of these stars on the sky and
  on their measured kinematics, we interpret them as belonging to the
  Magellanic Bridge. Our PMs seem to confirm N-body predictions for
  such a feature presented by \cite{schmidt20}, when extrapolated to
  the position of our field.}
\end{description}

Recently, \cite{omkumar20} reported on
the detection of a kinematically distinct sub-structure in front of
the SMC, which is particularly dominant between 2.5 and 5 degrees from
the SMC centre. This sub-structure has a PM very similar to the one we
measured for our secondary feature, and was interpreted by the authors
as the result of tidal stripping from the Magellanic Bridge.  Given that 
the kinematic properties of their sub-structure and our secondary 
component are fairly consistent, the two features might belong to
the same population of Magellanic Bridge stars, which we sampled at a
closer distance to the centre of the host galaxy. Future spectroscopic,
radial velocity measurements could help in confirming the possible
common origin of these two populations.

Because of their large distance and high density, it is challenging to
study the star clusters of the MCs, even for instruments like those
onboard the {\it Gaia} mission.  In this paper, we demonstrate that
the availability of multi-epoch {\it HST} observations makes it
possible to resolve and investigate the kinematics of the stellar
population in the fields of MC clusters. On one hand, this will
enhance the photometric investigation of elusive features in the CMD
of MC star clusters, which can be strongly contaminated by field
sources.  On the other hand, resolving the clusters' 3D kinematics
around the host galaxies will enable dynamical investigations that can
shed light on the assembly histories of the MCs (e.g., \citealt{piatti19}), 
in turn providing important constraints on the shape of the MCs' 
gravitational potentials.

\section*{Acknowledgements}
We warmly thank the anonymous referee for the constructive report 
that improved the quality of the paper.
Based on observations made with the NASA/ESA {\it Hubble Space Telescope},
obtained from the data archive at the Space Telescope Science
Institute.  STScI is operated by the Association of Universities for
Research in Astronomy, Inc. under NASA contract NAS 5-26555.  This
work has made use of data from the European Space Agency (ESA) mission
{\it Gaia} (http://www.cosmos.esa.int/gaia), processed by the {\it
  Gaia} Data Processing and Analysis Consortium (DPAC, {\tt
  http://www.cosmos.esa.int/web/gaia/dpac/consortium}). Funding for
the DPAC has been provided by national institutions, in particular the
institutions participating in the {\it Gaia} Multilateral Agreement.

\section*{Data availability statement}
The data underlying this article will be shared on reasonable request to the corresponding author.












\bsp	
\label{lastpage}
\end{document}